\documentclass[a4paper,10pt]{article}
\usepackage{amssymb}
\usepackage{amsmath}

\def \be {\begin{equation}}
\def \ee {\end{equation}}
\def \bea {\begin{eqnarray}}
\def \eea {\end{eqnarray}}
\def \nn {\nonumber}

\def \a {\alpha}
\def \b {\beta}

\def \d {\delta}
\def \eps {\epsilon}
\def \m {\mu}
\def \n {\nu}
\def \k {\kappa}

\def \s {\sigma}
\def \r {\rho}
\def \o {\omega}

\def \th {\theta}
\def \Th {\Theta}

\def \t {\tau}
\def \dag {\dagger}
\def \p {\partial}

\def\bd{\begin{document}}
\def\ed{\end{document}}
\def\nn{\nonumber}
\def\bea{\begin{eqnarray}}
\def\eea{\end{eqnarray}}
\let\bm=\bibitem
\let\la=\label

\def\N{{\cal N}}
\def\sst{\scriptscriptstyle}
\def\thetabar{\bar\theta}
\def\Tr{{\rm Tr}}
\def\one{\mbox{1 \kern-.59em {\rm l}}}

\def\a{\alpha}      \def\da{{\dot\alpha}}
\def\b{\beta}       \def\db{{\dot\beta}}
\def\c{\gamma}  \def\C{\Gamma}  \def\cdt{\dot\gamma}
\def\d{\delta}  \def\D{\Delta}  \def\ddt{\dot\delta}
\def\e{\epsilon}        \def\vare{\varepsilon}
\def\f{\phi}    \def\F{\Phi}    \def\vvf{\f}
\def\h{\eta}
\def\k{\kappa}
\def\l{\lambda} \def\L{\Lambda}
\def\m{\mu} \def\n{\nu}
\def\o{\omega}
\def\P{\Pi}
\def\r{\rho}
\def\s{\sigma}  \def\S{\Sigma}
\def\t{\tau}
\def\th{\theta} \def\Th{\Theta} \def\vth{\vartheta}
\def\X{\Xeta}
\def\z{\zeta}
\def\w{\wedge}
\def\u{\underline}
\def\hs{\hspace}

\def\cA{{\cal A}} \def\cB{{\cal B}} \def\cC{{\cal C}}
\def\cD{{\cal D}} \def\cE{{\cal E}} \def\cF{{\cal F}}
\def\cG{{\cal G}} \def\cH{{\cal H}} \def\cI{{\cal I}}
\def\cJ{{\cal J}} \def\cK{{\cal K}} \def\cL{{\cal L}}
\def\cM{{\cal M}} \def\cN{{\cal N}} \def\cO{{\cal O}}
\def\cP{{\cal P}} \def\cQ{{\cal Q}} \def\cR{{\cal R}}
\def\cS{{\cal S}} \def\cT{{\cal T}} \def\cU{{\cal U}}
\def\cV{{\cal V}} \def\cW{{\cal W}} \def\cX{{\cal X}}
\def\cY{{\cal Y}} \def\cZ{{\cal Z}}

\def\ua{\underline{\alpha}} \def\ubb{\underline{\beta}}
\def\ug{\underline{\gamma}}
\def\ub{\underline{\phantom{\alpha}}\!\!\!\beta}
\def\uc{\underline{\phantom{\alpha}}\!\!\!\gamma}
\def\um{\underline{\mu}} \def\un{\underline{\nu}}
\def\ud{\underline\delta}
\def\ue{\underline\epsilon}
\def\una{\underline a}\def\unA{\underline A}
\def\unb{\underline b}\def\unB{\underline B}
\def\unc{\underline c}\def\unC{\underline C}
\def\und{\underline d}\def\unD{\underline D}
\def\une{\underline e}\def\unE{\underline E}
\def\unf{\underline{\phantom{e}}\!\!\!\! f}\def\unF{\underline F}
\def\unm{\underline m}\def\unM{\underline M}
\def\unn{\underline n}\def\unN{\underline N}
\def\unp{\underline{\phantom{a}}\!\!\! p}\def\unP{\underline P}
\def\unq{\underline{\phantom{a}}\!\!\! q}
\def\unQ{\underline{\phantom{A}}\!\!\!\! Q}
\def\unH{\underline{H}}
\def\ul{\underline}

\def\As {{A \hspace{-6.4pt} \slash}\;}
\def\bs {{b \hspace{-6.4pt} \slash}\;}
\def\Ds {{D \hspace{-6.4pt} \slash}\;}
\def\ds {{\del \hspace{-6.4pt} \slash}\;}
\def\ss {{\s \hspace{-6.4pt} \slash}\;}
\def\ks {{ k \hspace{-6.4pt} \slash}\;}
\def\ps {{p \hspace{-6.4pt} \slash}\;}
\def\pas {{{p_1} \hspace{-6.4pt} \slash}\;}
\def\pbs {{{p_2} \hspace{-6.4pt} \slash}\;}

\def\Fh{\hat{F}}
\def\Vh{\hat{V}}
\def\Xh{\hat{X}}
\def\ah{\hat{a}}
\def\xh{\hat{x}}
\def\yh{\hat{y}}
\def\ph{\hat{p}}
\def\xih{\hat{\xi}}

\def\psit{\tilde{\psi}}
\def\Psit{\tilde{\Psi}}
\def\tht{\tilde{\th}}

\def\At{\tilde{A}}
\def\Qt{\tilde{Q}}
\def\Rt{\tilde{R}}
\def\Nt{\tilde{N}}

\def\at{\tilde{a}}
\def\st{\tilde{s}}
\def\ft{\tilde{f}}
\def\pt{\tilde{p}}
\def\qt{\tilde{q}}
\def\vt{\tilde{v}}
\def\nt{\tilde{n}}

\def\delb{\bar{\partial}}
\def\bz{\bar{z}}
\def\bD{\bar{D}}
\def\bB{\bar{B}}

\def\bk{{\bf k}}
\def\bl{{\bf l}}
\def\bp{{\bf p}}
\def\bq{{\bf q}}
\def\br{{\bf r}}
\def\bx{{\bf x}}
\def\by{{\bf y}}
\def\bR{{\bf R}}
\def\bV{{\bf V}}

\def\d{\delta}\def\D{\Delta}\def\ddt{\dot\delta}

\def\p{\partial} \def\del{\partial}
\def\xx{\times}
\def\uno{\mbox{1 \kern-.59em {\rm l}}}

\def\trp{^{\top}}
\def\inv{^{-1}}
\def\dag{{^{\dagger}}}

\def\pr{\prime}
\def\rar{\rightarrow}
\def\lar{\leftarrow}
\def\lrar{\leftrightarrow}

\begin{document}
\title{Spin-3  Topologically Massive Gravity}
\author{Bin Chen$^{1,2}$\footnote{Email:bchen01@pku.edu.cn}\hs{2ex}Jiang Long$^1$\footnote{Email:longjiang0301@gmail.com}\hs{2ex}Jun-bao Wu$^3$\footnote{Email:wujb@ihep.ac.cn}\\
\small{$^1$Department of Physics,
and State Key Laboratory of
Nuclear Physics and Technology,}\\
\small{Peking University, Beijing 100871, P.R. China}\\
\small{$^2$Center for High Energy Physics,
Peking University, Beijing 100871, P.R. China}\\
\small{$^3$Institute of High Energy Physics,
and Theoretical Physics Center for Science Facilities,}\\
\small{ Chinese Academy of Sciences,
Beijing 100049, P.R. China}}
\date{\today}
\maketitle


\begin{abstract}
In this paper, we study the spin-3 topologically massive gravity (TMG), paying special attention to its properties
at the chiral point. We propose an action
 describing the higher spin fields coupled to TMG. We discuss the traceless spin-3  fluctuations around the AdS$_3$ vacuum
 and find that there is an extra local massive mode, besides the left-moving and right-moving boundary massless modes.
 At the chiral point, such extra mode becomes massless and degenerates with the left-moving mode. We show that
 at the chiral point the  only degrees of freedom in the theory are the boundary right-moving graviton and spin-3 field. We conjecture that
 spin-3 chiral gravity with generalized Brown-Henneaux boundary condition is holographically dual to 2D chiral CFT with classical $W_3$ algebra and central charge  $c_R=3l/G$.
 \end{abstract}

\newpage






\section{Introduction}

The constructions of higher spin field theory has a long history \cite{Pauli}. Though in flat spacetime, a consistent interacting higher spin theory does not exist\cite{Aragone:1979hx},  it makes perfect sense in a spacetime background with a nonzero
cosmological constant \cite{Vasiliev:1980as, Vasiliev:1986td}. However when the dimension of the spacetime is higher than three, once we include one massless field
with spin higher than two, we must include an infinite number of massless fields with various higher spins and also other compensator fields. This fact makes the higher spin field theory much more difficult to deal with. For nice reviews on higher spin field theory, see \cite{Bekaert:2010hw}-\cite{Bouatta:2004kk}.


 In three dimensional case, the situation is quite different. It was found that there is no need to consider an infinite number of higher spin fields in order to have a consistent interaction. In $AdS_3$ case, not only the extra compensator fields vanish, the truncation to a field theory up to finite spin $N$ is also possible.
 In terms of  frame-like fields it was found long time ago that the higher spin field theory in $AdS_3$ could be rewritten as a Chern-Simons action\cite{Blencowe:1988gj,{Bergshoeff:1989ns}}. Very recently, it was realized in two remarkable papers\cite{Henneaux:2010xg,theisen} that the coupling of massless spin-3 field to the $AdS_3$ gravity could be described by a $SL(3)\times SL(3)$ Chern-Simons theory with opposite levels. Moreover in this Chern-Simons formulation, the asymptotic symmetry algebra of higher spin field theories\cite{Henneaux:2010xg, theisen} was shown to be a classical $W$-algebra with central charges  being the same as the ones obtained by Brown and Henneaux \cite{Brown:1986nw} in pure $AdS_3$ gravity. This opened a new window to study the higher spin $AdS_3$ gravity. The black hole solution with spin-3 hairs was constructed in \cite{Gutperle:2011kf,Ammon:2011nk}.
Partly based on \cite{Gaberdiel:2010ar}, a large $N$ duality for higher spin fields on $AdS_3$ was proposed\cite{Gaberdiel:2010pz}. The bulk theory includes massless higher spin fields and two extra scalars, while the boundary $CFT_2$ is a WZW coset model. This correspondence was checked in \cite{Gaberdiel:2011wb, Gaberdielnew,{Castro:2010ce}} and further developed in \cite{Ahn:2011pv,{Chang:2011mz}}. This duality can be thought as a kind of $AdS_3$ version of the Klebanov-Polyakov correspondence \cite{KP}, in which case  the higher spin field theories on $AdS_4$ is conjectured to be dual to $O(N)$ vector models in the large $N$ limit. Nontrivial supports to this conjectural duality came from the computations of the three-point correlation functions \cite{Yin1, Yin2, Yin3}. For an interesting derivation of higher spin gravity in AdS space from free bosonic field theory, see \cite{Douglas:2010rc} and \cite{Jevicki:2011ss}. String theory realization of spin-3 field was considered in \cite{Polyakov:2011sm}.

In the Chern-Simons formulation, it is clear that  both Einstein gravity \cite{Achucarro:1987vz, Witten:1988hc} and the higher spin field theories
 \cite{Blencowe:1988gj, Bergshoeff:1989ns} in $AdS_3$ have  no propagating mode. This is because that the Chern-Simons theory is a purely topological theory with no local degree of freedom, no matter what the gauge groups and the levels are. In the pure gravity case, in order to induce local degree of freedom, a gravitational Chern-Simons term with coefficient $\frac{1}{\m}$ had been included into the action. This gives us the so-called topologically massive gravity (TMG) theory\cite{Deser:1982vy, Deser:1981wh}. In TMG, there is one more local massive degree of freedom in $AdS_3$ background. However, it was found that the TMG theory is not stable for generic $\m$. It is only well-defined at the point $\m l=1$, the so-called chiral point. At this point, the massive graviton becomes massless, degenerates with the left-moving massless boundary graviton, and is just a pure gauge. It was conjectured that after imposing self-consistent Brown-Henneaux boundary condition the chiral gravity is holographically dual to a boundary chiral CFT with only right-moving degrees of freedom and central charge $c_R=3l/G$\cite{chiral,Maloney:2009ck}. At the chiral point, the level of left moving Chern-Simons theory vanishes.  Usually, the TMG theory is quite different from Chern-Simons gravity. But at the chiral point, the disappearance of the massive modes leads us to conjecture that the chiral gravity and holomorphic Chern-Simons gravity with only one  gauge group $SL(2,R)_R$ are equivalent in the following sense: when one impose the Brown-Henneaux boundary conditions for the fluctuations around the $AdS_3$ vacuum, these two theories have the same physical degrees of freedom, and may even have the same dynamics. The holomorphic nature of partition function of chiral gravity supports this equivalence\cite{Maloney:2009ck}(however see also \cite{Santamaria:2011cz}).


 In this paper, we would like to consider the higher spin field coupled to gravity in TMG theory. We will focus on the spin-3 case.  We will see that there indeed is a massive traceless spin-3 mode at generic value of $\m$. The conformal weight of such mode is $(\m l+2,\m l-1)$, which degenerates with the left-moving massless mode at the chiral point. The energies of left-moving massless mode and massive mode are vanishing at the chiral point, indicating that both of them are pure gauge. This fact suggests that the higher spin TMG theory becomes chiral at $\m l=1$, and is equivalent to a holomorphic Chern-Simons theory with one gauge group in the above sense. The asymptotic symmetries of such Chern-Simons theory is again just one copy of $W_3$ algebra with the central charge being the one of chiral gravity, taken generalized Brown-Henneaux boundary conditions \cite{Henneaux:2010xg,theisen}.

 In the next section,  by deforming the Chern-Simons action
with gauge group $SL(3)\times SL(3)$, we obtain the action for the
TMG coupled to the spin-3 field. With this action at hand, we
study several aspects of this theory. In section 3, we show that the traceless part of the spin-3 fluctuations around the $AdS_3$ vacuum
 satisfy a
third-order differential equation. We find that the equation could be decomposed into the product of three first-order differential
equations, describing two massless and one massive modes. As the equation of all the modes could be changed into a second order
differential equation, which could be rewritten in terms of $SL(2,R)$ Casimir, we
 compute the conformal weight of the
fluctuations and obtain the explicit solution of fluctuations with the
highest conformal weight in section 4. We also discuss the logarithmic mode, besides the massless and massive modes. Such log modes could be truncated by imposing generalized Brown-Henneaux boundary conditions. In section 5, we calculate the energies of traceless spin-3
fluctuations and show that only at the chiral point, there are no
modes with negative energy, and at this point, both massive mode and left
moving mode have zero energy.
We end with conclusions and some discussions in section 6.  We conjecture that
the chiral spin-3 gravity is holographically dual to a chiral 2D CFT with $W_3$ symmetry, with the generalized Brown-Henneaux boundary conditions.

\section{Action and equations of motion}

In this section, we propose an action describing the spin-3 field coupled to 3D TMG theory with a negative cosmological constant. As we will work in the frame-like formulation, we give a brief review of first order formalism of TMG, and discuss the possible relation with a Chern-Simons theory. Next we show how to incorporate the spin-3 field in our formulation.

\subsection{First order formulation of TMG}

In first order formalism, TMG with a negative cosmological constant $\Lambda=-l^{-2}$ is described by the action
\be
S_{\mbox{\tiny TMG}}=  \frac1{8\pi G}\int\big(e^a\w R_a+\frac1{6l^2}\epsilon_{abc}e^a\w e^b \w
e^c\big)-\frac1{16\pi G\mu}\int \big({\cal L}_{\mbox{\tiny CS}}+\beta^a\w T_a\big),
\ee
where
\bea
R_a&=&d\omega_a+\frac12\epsilon_{abc}(\omega^b\w \omega^c+\frac{e^b\w
e^c}{l^2} ) \nn\\
T^a&=&de^a+\epsilon^{abc}\omega_b\w
e_c,\nn\\
{\cal L}_{\mbox{\tiny CS}}&=&\omega^a\w d\omega_a+\frac13
\epsilon_{abc}\omega^a\w\omega^b\w\omega^c.
\eea
The field $\b^a$ is just a Lagrangian multiplier, imposing the torsion free condition
such that the above action is equivalent to the action in terms of Christoffel symbol\cite{Deser:1991qk,{Carlip:1991zm}}.

It would be illuminating to rewrite the above action in a form relating to Chern-Simons gravity with gauge
group $SL(2,R)\times SL(2,R)$. In order
to do so, we need to combine the frame-like fields and the spin connections into two gauge fields:
\be
A=(\omega^a_\mu+\frac1l e^a_\mu)J_a dx^\mu,\hs{3ex}
\tilde
A=(\omega^a_\mu-\frac1l e^a_\mu)J_adx^\mu.\ee
Here $l$ is the AdS radius and $J_a$'s are the generators of $SL(2,R)$. Then the action of TMG  could be rewritten as
\be
S_{\mbox{\tiny TMG}}=\left(1-\frac1{\mu
l}\right)S_{CS}[A]-\left(1+\frac1{\mu l}\right)S_{CS}[\tilde
A]-\frac{k}{4\pi\mu l}\int\left(\tilde\beta^a\w T_a\right), \ee
where
\be S_{CS}[A]=\frac
k{4\pi}\int{\rm Tr}(A\w dA+\frac23 A\w A\w A),\ee
with
\be
k=\frac{l}{4G}, \hs{3ex}\tilde\b^a=\b^a-\frac{e^a}{l^2}.\ee

Note that the solutions of usual AdS$_3$ gravity are automatically
the solutions of TMG theory. Especially, the Einstein metric
solution leads to $\tilde\b^a=0$. Therefore it seems that the TMG in
AdS$_3$ background is equivalent to Chern-Simons
theory\cite{CDWW1,{CDWW2}}. However, one should notice that TMG has
one local degree of freedom whereas CS theory has none. In this
case, the fluctuations around the AdS$_3$ background satisfy a third
order differential equation, reflecting the fact that the
gravitational Chern-Simons term includes higher derivative terms,
and more importantly leads to a local massive mode. This suggests
that the TMG theory is very different from the Chern-Simons gravity,
which could be just a topological theory. The right way is just to
keep the Lagrangian multiplier to impose the torsion free
condition.

Nevertheless the story could be different at the chiral point\cite{chiral}. The conformal weight of the massive graviton
in AdS$_3$ is $\frac{1}{2}(3+\m l, -1+\m l)$, which becomes degenerate with the one of the left-moving massless graviton. Thus there is no local degree of freedom anymore so that chiral gravity has the same degree of freedom as a holomorphic Chern-Simons gravity, after imposing the Brown-Henneaux boundary condition to truncate the log modes. Moreover  the central charge of asymptotic symmetry group become $c_L=0, c_R=\frac{3l}{G}$. On the other hand, the asymptotic symmetry of Chern-Simons gravity suggests that the central charges are only related to the levels
\be
c_L= (1-\frac1{\mu l})\frac{3l}{2G}, \hs{3ex}c_R=(1+\frac1{\mu l})\frac{3l}{2G},
\ee
which reproduce precisely the central charges of chiral gravity. More importantly, the partition function of chiral gravity is holomorphic. All these facts support that the chiral gravity is equivalent to a holomorphic Chern-Simons gravity in the sense we described in the introduction.

\subsection{Action of spin-3 TMG}

We may generalize the above construction to include the spin-3 field.
First of all, we need to replace the gauge group $SL(2, R)$ with $SL(3,R)$.
The $SL(3, R)$ group has the generators  $J_a, T_{ab} (a, b=1, 2,
3)$ with $T_{ab}$ being symmetric and traceless. They satisfy the
following commutation relations: \be [J_a, J_b]=\eps_{abc}J^c,\hs{3ex}
[J_a, T_{bc}]=\eps^d_{\,\,\,a(b}T_{c)d},\ee \be [T_{ab},
T_{cd}]=\sigma(\eta_{a(c}\eps_{d)be}+\eta_{b(c}\eps_{d)ae})J^e. \ee
For the inner products of the generators, we have \bea {\rm
Tr}(J_aJ_b)=\frac12\eta_{ab}, \hs{3ex} {\rm Tr}(J_aT_{bc})=0, \\ {\rm
Tr}(T_{ab}T_{cd})=-\frac\sigma2(\eta_{a(c}\eta_{d)b}-\frac23\eta_{ab}\eta_{cd}).\eea
There is actually a free parameter $\sigma$ in the above algebraic relation. In this paper, we focus on the case $\sigma <0$. Up to an overall factor, the right handed side of the last equation
can be determined by that it should be symmetric and traceless with
respect to $ab$ and $cd$, respectively. And the overall factor is
fixed by that we have:\be {\rm Tr}([J_a, T_{bc}]T_{de})={\rm
Tr}(J_a[T_{bc}, T_{de}]).\ee

As in \cite{theisen},  we combine the
vielbein-like fields and the connections of spin-2 and spin-3 into two
gauge potentials $A, \tilde A$£º \bea
A&=&((\omega^a_\mu+\frac1le^a_\mu)J_a+(\omega^{ab}_\mu+\frac1le^{ab}_\mu)T_{ab})dx^\mu,\\
\tilde
A&=&((\omega^a_\mu-\frac1le^a_\mu)J_a+(\omega^{ab}_\mu-\frac1le^{ab}_\mu)T_{ab})dx^\mu.\eea
 It was shown \cite{theisen} that the
action \be S=S_{CS}[A]-S_{CS}[\tilde A],\label{CS}\ee  gives the correct
theory for spin-3 field coupled to AdS$_3$ gravity.

To study the topologically massive gravity coupled with spin-3 fields,
we now propose the following action: \be S=\left(1-\frac1{\mu
l}\right)S_{CS}[A]-\left(1+\frac1{\mu l}\right)S_{CS}[\tilde
A]-\frac{k}{4\pi\mu l}\int\left(\tilde\beta^a\w T_a-2\sigma
\tilde{\beta}^{ab}\w T_{ab}\right). \label{CS form}\ee Here  the last two terms are introduced to impose the
torsion free conditions.
The torsions are still the same as the ones in \cite{theisen}:
 \bea T^a&=&de^a+\epsilon^{abc}\omega_b\w
e_c-4\sigma\epsilon^{abc}e_{bd}\w
\omega_c^{\,\,d},\\
T^{ab}&=&de^{ab}+\epsilon^{cd(a|}\omega_c\w
e_d^{\,\,|b)}+\epsilon^{cd(a|} e_c\w \omega_d^{\,\,|b)},\eea
obtained from the equations of motion there from the action eq.~(\ref{CS}).
Note that the torsion $T^a$ for veilbein gets modified by the spin-3 field and the torsion $T^{ab}$ is for spin-3 field.

In terms of the frame-like field and connection, the action (\ref{CS form}) could be written
in a more familiar form:
\bea S&=&\frac1{8\pi G}\int\big( e^a\w
d\omega_a+\frac12 \epsilon_{abc}e^a\w
\omega^b\w\omega^c+\frac1{6l^2}\epsilon_{abc}e^a\w e^b \w
e^c\nn\\
&&-2\sigma e^{ab}\w
d\omega_{ab}-2\sigma\epsilon_{abc}e^a\w\omega^{bd}\w\omega^c_{\,\,d}-2\sigma
e^{ab}\w\epsilon_{(a|cd}\omega^c\w \omega^{\,\,\,\,d}_{|b)}
\nn\\
&&-\frac{2\sigma}{l^2}\epsilon_{abc}e^a\w e^{bd}\w e^c_{\,\,d}\big)
-\frac1{16\pi G\mu}\int \big( \omega^a\w d\omega_a+\frac13
\epsilon_{abc}\omega^a\w\omega^b\w\omega^c\nn\\
&& -2\sigma\omega^{ab}\w
d\omega_{ab}-4\sigma\epsilon_{abc}\omega^a\w\omega^{bd}\w\omega^c_{\,\,d}
+\beta^a\w T_a-2\sigma \beta^{ab}\w T_{ab}\big),\label{action} \eea
where
\be
\b^{ab}=\tilde\b^{ab}+\frac{e^{ab}}{l^2}
\ee
could be taken as an independent field. The first part of the action, proportional to $1/8\pi G$, is the same as the
one of pure spin-3 AdS$_3$ gravity found in \cite{theisen}. The two terms proportional to $T_a,T_{ab}$ are just to impose the torsion
free conditions. The remaining parts are the spin-3 generalization of gravitational Chern-Simons term.

The equations of motion from this action are: \bea T^a=0, \hs{3ex}
T^{ab}=0,\eea
which are the torsion free conditions, and
\bea && R_a-\frac1{2\mu
}(d\beta_a+\epsilon_{abc}\beta^b\w
\omega^c-2\sigma\epsilon_{(c|da}\beta^{bc}\w \omega^d_{\,\,|b)})=0,\\
&& R_a+\frac12\epsilon_{abc}\left[\beta^b\w e^c-\frac{e^b\w
e^c}{l^2}+4\sigma\left(\frac{e^{bd}\w e^c_{\,\,d}}{l^2}-e^{bd}\w
\beta^c_{\,\,d}\right)\right]=0,\\
&& R_{ab}-\frac1{2\mu}\left(d\beta_{ab}+\epsilon_{cd(a|}\beta^c\w
\omega^d_{\,\,|b)}+\epsilon_{cd(a|}\omega^c\w
\beta^d_{\,\,|b)}\right)=0,\\
&& R_{ab}+\frac12\left(\epsilon_{cd(a|}\beta^c\w
e^d_{\,\,|b)}+\epsilon_{cd(a|}e^c\w
\beta^d_{\,\,|b)}\right)-\frac1{l^2}\epsilon_{cd(a|}e^c\w
e^d_{\,\,|b)}=0, \eea where \bea
R_a&=&d\omega_a+\frac12\epsilon_{abc}(\omega^b\w \omega^c+\frac{e^b\w
e^c}{l^2} )-2\sigma\epsilon_{abc}(\omega_{bd}\w
\omega_c^{\,\,d}+\frac{e_{bd}\w e_c^{\,\,d}}{l^2}),\\
R_{ab}&=&d\omega_{ab}+\epsilon_{cd(a|}\omega^c\w
\omega^d_{\,\,|b)}+\frac{1}{l^2}\epsilon_{cd(a|}e^c\w
e^d_{\,\,|b)}.\eea

The equations are different from the pure gravity case, with contributions from $\b^a,\b^{ab}$ terms.
However, notice that when
\be
\b^a=\frac{e^a}{l^2}, \hs{3ex} \b^{ab}=\frac{e^{ab}}{l^2},
\ee
the extra terms vanish due to the torsion free conditions. This suggests that the solutions of pure gravity coupled to
spin-3 field theory proposed in \cite{theisen} remain the solutions of above equations of motion. But it could be possible that there
exist more solutions. Actually if we just consider the case with vanishing spin-3 fields, then the above equations allow the solutions
besides the Einstein-type metric. Two classes of such solutions are the warped spacetimes and null spacetimes\cite{Andy08,{Bin10selfdual}}.
In this paper, we just focus on the AdS$_3$ spacetime or its discrete quotient--BTZ black hole\cite{Banados:1992wn}.

\section{Spin-3 fluctuations}

Now we consider the fluctuations around a given fixed background
with background fields $\bar{e}^a, \bar{\omega}^a, \bar{\beta}^a,
\bar{e}^{ab}, \bar{\omega}^{ab}, \bar{\beta}^{ab}$. For simplicity, we only consider
the backgrounds with $\bar{e}^{ab}=0$, then we
have $\bar{\omega}^{ab}=\bar{\beta}^{ab}=0$. We still denote the
fluctuations of the fields as $e^a, \cdots$. To the leading order,
we have  the following equation of motions of the fluctuations: \bea
de^a+\epsilon^{abc}\bar\omega_b\w
e_c+\epsilon^{abc}\omega_b\w \bar{e}_c=0,\\
d\omega_a+\epsilon_{abc}(\bar\omega^b\w
\omega^c+\frac{\bar{e}^b\w e^c}{l^2})-\frac1{2\mu
}(d\beta_a+\epsilon_{abc}\bar\beta^b\w
\omega^c+\epsilon_{abc}\beta^b\w \bar\omega^c)=0,\\
d\omega_a+\epsilon_{abc}(\bar\omega^b\w
\omega^c+\frac{\bar{e}^b\w
e^c}{l^2})+\frac12\epsilon_{abc}\left[\bar\beta^b\w e^c+\beta^b\w
\bar{e}^c-\frac{2}{l^2}\bar{e}^b\w e^c\right]=0,\\
de^{ab}+\epsilon^{cd(a|}\bar{\omega}_c\w
e_d^{\,\,|b)}+\epsilon^{cd(a|} \bar{e}_c\w
\omega_d^{\,\,|b)}=0,\label{torsion2}\\
R_{ab}-\frac1{2\mu}\left(d\beta_{ab}+\epsilon_{cd(a|}\bar\beta^c\w
\omega^d_{\,\,|b)}+\epsilon_{cd(a|}\bar\omega^c\w
\beta^d_{\,\,|b)}\right)=0,\label{eom2}\\
R_{ab}+\frac12\left(\epsilon_{cd(a|}\bar\beta^c\w
e^d_{\,\,|b)}+\epsilon_{cd(a|}\bar{e}^c\w
\beta^d_{\,\,|b)}\right)-\frac1{l^2}\epsilon_{cd(a|}\bar{e}^c\w
e^d_{\,\,|b)}=0, \label{eom1}\eea where  \bea
R_{ab}=d\omega_{ab}+\epsilon_{cd(a|}\bar\omega^c\w
\omega^d_{\,\,|b)}+\frac{1}{l^2}\epsilon_{cd(a|}\bar{e}^c\w
e^d_{\,\,|b)}.\eea

To the leading order, the
fluctuations of the spin-3 fields decouple from
the gravitons. And the latter was studied in detail around the
$AdS_3$ background in \cite{chiral}. Actually, one can show that in our formalism, the fluctuations of gravitons $h_{\mu\nu}$ satisfy the following equation
\be
(\Box+\frac{2}{l^2})h^{\rho}_{\sigma}+\frac{1}{\mu}\epsilon^{\rho\mu\nu}\nabla_{\mu}(\Box+\frac{2}{l^2})h_{\nu\sigma}=0,\label{spin2eq}
\ee
where the transverse and traceless gauge conditions $\nabla^{\mu}h_{\mu\nu}=0$ and $h^{\mu}_{\mu}=0$ have been chosen. It is exactly the same as the one in  \cite{chiral}.

Next we focus on the
fluctuations of spin-3 fields. Let us define
$\Phi_{\mu\nu\lambda}=e_{\mu ab}\bar{e}^a_\nu \bar{e}^b_\lambda$.
From $T^{ab}=0$, we obtain that: \bea
\omega^{ab}_\mu&=&\frac13\eta^{ab}\epsilon^{\nu\rho\lambda}\nabla_\nu
\Phi_{\rho\lambda\mu}+\frac12
\epsilon_{\mu}^{\ \nu\rho}\nabla_\nu\Phi_{\rho\alpha\beta}\bar{e}^{\alpha
a}\bar{e}^{\beta b}- \frac12 \bar{e}^{(a|}_\nu
\bar{e}^{|b)\phi}\epsilon^{\nu\rho\lambda}\nabla_\rho
\Phi_{\lambda\phi\mu}. \nn\eea

Actually
Eq.~(\ref{eom1}) can be rewritten as \bea
R_{ab}+\frac12\left(\epsilon_{cd(a|}\tilde{\bar\beta}^c\w
e^d_{\,\,|b)}+\epsilon_{cd(a|}\bar{e}^c\w
\tilde\beta^d_{\,\,|b)}\right)=0,\eea from which we get \be
\tilde\beta_\mu^{ab}=\frac23\eta^{ab}\bar{e}_{\mu c}Q_d^{\,\, dc}+\bar{e}_{\mu
c}(Q^{cab}-Q^{abc}-Q^{bac}),\label{tbeta}\ee where \be Q^{\mu}
_{\ ab}=\epsilon^{\mu\nu\lambda}(R_{ab\nu\lambda}+\frac14\epsilon_{cd(a|}\tilde{\bar\beta}^c_{[\nu}e_{\lambda]\,\,\,|b)}^{\,\,d}).\ee
Now we are considering the background $AdS_3$ or BTZ black hole, then
we have $\tilde{\bar\beta}^a=0$. From the definition of $R^{ab}$ we
find that in this case: \bea Q^{\rho ab}\bar{e}^\alpha_a
\bar{e}^\beta_b&=&\epsilon^{\rho\mu\nu}R^{ab}_{\mu\nu}\bar{e}^\alpha_a
\bar{e}^\beta_b\nn\\&=&\frac13
g^{\a\b}\eps^{\rho\mu\nu}\eps^{\sigma\lambda\kappa}\nabla_\mu\nabla_\lambda
\Phi_{\kappa\sigma\nu}+\frac12\Box\Phi^{\rho\alpha\beta}-\frac12\nabla_\mu\nabla^\rho\Phi^{\mu\alpha\beta}\nn\\
&&-\frac12\eps^{\lambda\kappa(\alpha|}\eps^{\rho\mu\nu}\nabla_\mu\nabla_\lambda
\Phi_{\kappa\nu}^{\,\,\,\,|\beta)}+\frac1{l^2}(\Phi^{(\alpha\beta)\rho}-g^{\rho(\alpha|}\Phi_\mu^{\,\,\,\mu|\beta)})\nn\\&=&-\frac13
g^{\a\b}(\Box
\Phi_\nu^{\,\,\,\nu\rho}+\nabla_\mu\nabla_\nu\Phi^{\rho\mu\nu}-\nabla_\mu\nabla_\nu\Phi^{\mu\nu\rho}-\nabla_\mu\nabla^\rho\Phi_{\nu}^{\,\,\,\nu\mu})
\nn\\
&&-\frac12\Box\Phi^{\rho\alpha\beta}+\frac12\nabla_\mu\nabla^\rho\Phi^{\mu\alpha\beta}
+\frac12(g^{(\alpha|\rho}\Box\Phi_\nu^{\,\,\,\nu|\beta)}-g^{(\alpha|\rho}\nabla_\mu\nabla_\nu
\Phi^{\mu\nu|\beta)}\nn\\
&&+\nabla^{(\a|}
\nabla_\mu\Phi^{\rho\mu|\beta)}-\nabla^{(\a|}
\nabla^\rho\Phi^{\,\,\,\mu|\beta)}_\mu)+\frac1{l^2}(\Phi^{(\alpha\beta)\rho}-g^{\rho(\alpha|}\Phi_\mu^{\,\,\,\mu|\beta)}).\nn\eea
As $\Phi^{\rho\mu\nu}$ includes hooked $\{2, 1\}$ component, we can use the
gauge transformation to get rid of them and then $\Phi^{\rho\mu\nu}$ is totally symmetric. If we further impose the
following condition to read the transverse traceless modes: \be \Phi_\mu^{\,\,\,\mu\rho}=0, \hs{3ex}
\nabla^\mu\Phi_{\mu\nu\rho}=0,\ee we get that \bea Q^{\rho
ab}\bar{e}^\alpha_a
\bar{e}^\beta_b=-\frac12\Box\Phi^{\rho\a\b}+\frac12\nabla_\mu\nabla^\rho\Phi^{\mu\a\b}+\frac2{l^2}\Phi^{\rho\a\b}
=-\frac12\Box\Phi^{\rho\a\b}.\label{q} \eea This result and
Eq.~(\ref{tbeta}) lead to \be
\tilde\beta^{\mu\a\b}=\frac12\Box\Phi^{\mu\a\b}.\label{beta} \ee In
such backgrounds, 
Eq.~(\ref{eom2}) gives: \be\
R^{ab}-\frac1{2\mu}(d\tilde\beta^{ab}+\eps_{cd(a|}\bar{\omega}^c\w\tilde\beta^d_{\,\,\,|b)})=0,\ee
which leads to \be
Q^{\rho\a\b}-\frac1{2\mu}\eps^{\rho\mu\nu}\nabla_\mu\tilde\beta_{\nu}^{\,\,\,\a\b}=0.\ee
Using Eqs.~(\ref{q}-\ref{beta}), we finally obtain \be \Box
\Phi^{\rho\a\b}+\frac1{2\mu}\eps^{\rho\mu\nu}\nabla_\mu \Box
\Phi_\nu^{\,\,\,\a\b}=0.\label{spin 3 eq}\ee
Compared to the equation (\ref{spin2eq}) of the graviton fluctuation, the spin-3 fluctuations satisfy a third order differential equation as well.

\section{Conformal weight of spin-3 fluctuations}

In $AdS_3$ or $BTZ$ background, we can rewrite the spin-3 equation as
\be
\Box(
\Phi^{\rho\a\b}+\frac1{2\mu}\eps^{\rho\mu\nu}\nabla_\mu
\Phi_\nu^{\,\,\,\a\b})=0,
\ee
where we have used the property
$R_{\rho\sigma\mu\nu}=-\frac{1}{l^2}(g_{\rho\mu}g_{\sigma\nu}-g_{\rho\nu}g_{\sigma\mu})$.
This third order differential equation could be decomposed into three first-order differential equations, each corresponding to different degrees of freedom.  The massive degree of freedom satisfies a first order equation
\be
\eps^{\rho\mu\nu}\nabla_\mu
\Phi_\nu^{(M)\,\,\,\a\b}=-2\mu\Phi^{(M)\rho\a\b}.
\ee
Both the left mover and right mover are massless and satisfy respectively
\be
\eps^{\rho\mu\nu}\nabla_\mu
\Phi_\nu^{(L)\,\,\,\a\b}=-\frac{2}{l}\Phi^{(L)\rho\a\b},\hs{3ex}\eps^{\rho\mu\nu}\nabla_\mu
\Phi_\nu^{(R)\,\,\,\a\b}=\frac{2}{l}\Phi^{(R)\rho\a\b}.
\ee
The three equations share the same structure, which could be denoted simply as
\be
\eps^{\rho\mu\nu}\nabla_\mu
\Phi_\nu^{(A)\,\,\,\a\b}=m_{A}\Phi^{(A)\rho\a\b},
\ee
Where $A$ can be $M,L,R$ and $m_{A}=-2\mu,-\frac{2}{l},\frac{2}{l}$ correspondingly.
One can derive the second order equation it satisfies as
\be
\Box\Phi^{(A)}_{\rho\mu\nu}=(m_{A}^2-\frac{4}{l^2})\Phi^{(A)}_{\rho\mu\nu}.\label{second order}
\ee

In the global coordinates the metric of $AdS_3$ is: \be ds^2=l^2
(-\cosh^2\rho d\tau^2+\sinh^2\rho d\phi^2+d\rho^2). \ee It has the
isometry group $SL(2, R)_L\times SL(2, R)_R$. By defining $u\equiv
\tau+\phi, v\equiv \tau-\phi$, the generators of
$SL(2, R)_L$ can be written as: \bea V_0&=&i\partial_u, \\
V_{-1}&=&ie^{-iu}\left(\frac{\cosh2\rho}{\sinh2\rho}\partial_u-\frac{1}{\sinh2\rho}\partial_v+\frac{i}2\partial_\rho\right),\\
V_{1}&=&ie^{iu}\left(\frac{\cosh2\rho}{\sinh2\rho}\partial_u-\frac{1}{\sinh2\rho}\partial_v-\frac{i}2\partial_\rho\right),\eea
satisfying \be [V_i,V_j]=(i-j)V_{i+j}. \ee By exchanging $u$ and $v$
in the above equations, we can get the generators $\bar{V}_0,
\bar{V}_{-1}, \bar{V}_1$ of $SL(2, R)_R$.

 Let us define \be
 \mathcal{L}^2=\mathcal{L}_{V_0}\mathcal{L}_{V_0}-\frac{1}{2}(\mathcal{L}_{V_1}\mathcal{L}_{V_{-1}}+\mathcal{L}_{V_{-1}}\mathcal{L}_{V_{1}}), \ee and similarly for $\bar{\mathcal{L}}^2$. Then we can rewrite Eq.~(\ref{second order}) as
\be
\mathcal{L}^2\Phi^{(A)}_{\rho\mu\nu}=\frac{(m_{A}l)^2-6(m_{A}l)+8}{4}\Phi^{(A)}_{\rho\mu\nu},\label{Lie derivative equation 1}
\ee
or
\be
\bar{\mathcal{L}}^2\Phi^{(A)}_{\rho\mu\nu}=\frac{(m_{A}l)^2+6(m_{A}l)+8}{4}\Phi^{(A)}_{\rho\mu\nu}.\label{Lie derivative equation 2}
\ee
Considering the highest weight state with conformal weight $(h^{(A)},\bar{h}^{(A)})$,
\bea {\cal L}_{V_1}\Phi^{(A)}_{\mu\nu\lambda}&=&{\cal
L}_{\bar{V}_1}\Phi^{(A)}_{\mu\nu\lambda}=0,\\ {\cal
L}_{V_0}\Phi^{(A)}_{\mu\nu\lambda}&=&h^{(A)}\Phi^{(A)}_{\mu\nu\lambda},\hs{3ex} {\cal
L}_{\bar{V}_0}\Phi^{(A)}_{\mu\nu\lambda}=\bar{h}^{(A)}\Phi^{(A)}_{\mu\nu\lambda}, \eea
then we have
\be
\mathcal{L}^2\Phi^{(A)}_{\rho\mu\nu}=(h^{(A)2}-h^{(A)})\Phi^{(A)}_{\rho\mu\nu},
\ee
and similar equation for $\bar{h^{(A)}}$. Comparing these two equations with Eq.~(\ref{Lie derivative equation 1}) and Eq.(\ref{Lie derivative equation 2}), we find the conformal weights as
\be
h^{(A)}=\frac{1\pm(m_A l-3)}{2}, \hs{3ex} \bar{h}^{(A)}=\frac{1\pm(m_A l+3)}{2}.
\ee
As the conformal weight should be positive, then the suitable choice of the conformal weights is
\bea
h^{(M)}=\mu l+2,&&\bar{h}^{(M)}=\mu l-1;\\
h^{(L)}=3,& &\bar{h}^{(L)}=0;\\
h^{(R)}=0,& &\bar{h}^{(R)}=3.
\eea
where we have assumed $\mu l\ge1$.
Note that for the chiral gravity at $\mu l=1$, we have $h^{(M)}=3$ and $\bar{h}^{(M)}=0$, which
degenerates with the left-moving massless spin-3 fluctuations.

Actually, we can work out the explicit solution for the spin-3 fluctuations of the
highest conformal weight. From the Killing symmetry of the background, we may make ansatz
\be
\Phi_{\mu\nu\lambda}=e^{-ihu-i\bar{h}v}F_{\mu\nu\lambda},\label{solution1}\ee
where all the components of $F_{\mu\nu\lambda}$ could be characterized by one undetermined function $\varphi$
\bea
F_{\tau\tau\tau}=\pm\varphi,\hs{2ex}
F_{\phi\phi\phi}=\varphi,\hs{2ex}
F_{\rho\rho\rho}=\mp \frac{i\varphi}{\sinh^3\rho\cosh^3\rho},\hs{2ex}
F_{\tau\tau\phi}=\varphi,\\
F_{\tau\tau\rho}=\pm\frac{i\varphi}{\sinh\rho\cosh\rho},\hs{2ex}
F_{\phi\phi\tau}=\pm\varphi,\hs{2ex}
F_{\phi\phi\rho}=\pm \frac{ i \varphi}{\sinh\rho\cosh\rho},\\
F_{\rho\rho\tau}=\mp \frac{\varphi}{\sinh^2\rho\cosh^2\rho},\hs{2ex}
F_{\rho\rho\phi}=-\frac{\varphi}{\sinh^2\rho\cosh^2\rho},\hs{2ex}
F_{\tau\rho\phi}=\frac{i\varphi}{\sinh\rho\cosh\rho}, \eea
The function $\varphi$ satisfies: \be \partial_\rho
\varphi+\frac{(h+\bar{h})\sinh^2\rho-3\cosh^2\rho}{\sinh\rho\cosh\rho}\varphi=0,
\ee whose solution is: \be
\varphi=C(\cosh\rho)^{-(h+\bar{h})}\sinh^3\rho\label{solution2}\ee with  a constant
$C$.

Besides the massless and massive modes,  there is actually a logarithmic spin-3 mode  at the chiral point, just like its spin-2
counterpart\cite{CDWW1,CDWW2,Grumiller:2008qz}. This is feasible because the
equation of fluctuations (\ref{spin 3 eq}) is a third order differential equation. The log mode can be constructed in a way similar to its spin-2 counterpart\cite{Grumiller:2008qz}
\be
\Phi^{(Log)}_{\mu\nu\sigma}\equiv\lim_{\mu l\to1}\frac{\Phi^{(M)}_{\mu\nu\sigma}-\Phi^{(L)}_{\mu\nu\sigma}}{\mu l-1}.
\ee
We define a function $A(\tau,\rho)$ as
\be
A(\tau,\rho)\equiv-2(i\tau+\ln\cosh\rho),
\ee
then the log mode can be written as
\be
\Phi^{(Log)}_{\mu\nu\sigma}=A(\tau,\rho)\Phi^{(L)}_{\mu\nu\sigma}.
\ee
This new mode grows linearly in time, and grows logarithmically in the radial coordinate $\rho$. By using the fact that \be
\mathcal{L}_{V_0}A=\bar{\mathcal{L}}_{\bar{V}_0}A=1;\hs{3ex}\mathcal{L}_{V_1}A=\bar{\mathcal{L}}_{\bar{V}_1}A=0,
\ee
One can show that
\be
\Box\Phi^{(Log)}_{\mu\nu\sigma}\propto\Phi^{(L)}_{\mu\nu\sigma}.
\ee
Hence the log mode satisfies the classical equations of motion Eq.(\ref{spin 3 eq}).

The existence of the log graviton mode in TMG brought much debates on the stability and the chiral nature of the theory. It turns out that one has to impose appropriate boundary conditions on the metric fluctuations\cite{Maloney:2009ck}. If one impose the generalized Brown-Henneaux boundary conditions \cite{Henneaux:2010xg,theisen}, such mode disappears and the only physical modes are right-moving boundary gravitons, then the theory is chiral. On the other hand, one may relax the boundary condition to allow the log mode, whose presence breaks the chiral nature of the theory. Nevertheless, such relaxed boundary condition is well-defined, leading to finite conserved charge. It was conjectured that the log gravity could be dual to a logarithmic CFT\cite{Grumiller:2008qz,Maloney:2009ck}. In this paper, we will not discuss such log mode, and just focus on the generalized Brown-Henneaux boundary conditions.

\section{Energies of spin-3 fluctuations}

As we show, for generic value of $\m l$, there are three kinds of spin-3 fluctuations, two of them being massless and the other one being massive. The fluctuation could be written as \be
\Phi_{\m_1\m_2\m_3}=\Phi^L_{\m_1\m_2\m_3}+\Phi^R_{\m_1\m_2\m_3}+\Phi^M_{\m_1\m_2\m_3}
\ee
where the subscript $L$ represent the (3,0) primary and their descendants, $R$ represent
the (0,3) primary and their descendants, and $M$ represents the ($\m l+2, \m l-1$) primary and their descendants.

Up to a positive normalization constant $C$, the free action of the spin-3 fluctuations is of the form
\be
S_2=\frac{1}{64\pi G}\int d^3x \sqrt{-g}\left\{-\bar{\nabla}^\l \Phi^{\m_1\m_2\m_3}\bar{\nabla}_\l \Phi_{\m_1\m_2\m_3}-\frac{1}{2\m}\bar{\nabla}_\a \Phi_{\m_1\m_2\m_3}\epsilon^{\m_1\a\b}\Box\Phi_{\b\m_2\m_3}\right\}.
\nn\ee
The conjugate momentum is
\be
\Pi^{(1)\m_1\m_2\m_3}=-\frac{\sqrt{-g}}{64\pi G}\left(\bar{\nabla}^0(2\Phi^{\m_1\m_2\m_3}+
\frac{1}{2\m}\epsilon^{\m_1\a}_{~~~\b}\bar{\nabla}_\a\Phi^{\b\m_2\m_3})-\frac{1}{2\m}
\epsilon_{\b}^{~0\m_1}\Box\Phi^{\b\m_2\m_3}\right).\nn
\ee
Using the equations of motion for different modes, we find
\bea
\Pi^{(1)\m_1\m_2\m_3}_L&=&-\frac{\sqrt{-g}}{64\pi G}(2-\frac{1}{\mu l})\bar{\nabla}^0\Phi_L^{\m_1\m_2\m_3},\\
\Pi^{(1)\m_1\m_2\m_3}_R&=&-\frac{\sqrt{-g}}{64\pi G}(2+\frac{1}{\mu l})\bar{\nabla}^0\Phi_R^{\m_1\m_2\m_3},\\
\Pi^{(1)\m_1\m_2\m_3}_M&=&\frac{\sqrt{-g}}{64\pi G}\left(-\bar{\nabla}^0\Phi_M^{\m_1\m_2\m_3}+\frac{2}{\m}(\m^2-\frac{1}{l^2})\epsilon_{\b}^{~0\m_1}\Phi_M^{\b\m_2\m_3}\right).
\eea

As there are three time derivatives in the action, we need to use the Ostrogradsky method to define the Hamiltonian. Taking $K_{\m_1\m_2\m_3}\equiv \bar{\nabla}_0\Phi_{\m_1\m_2\m_3}$ as a canonical variable, we find its conjugate momentum
\be
\Pi^{(2)\m_1\m_2\m_3}=-\frac{\sqrt{-g}g^{00}}{128\pi G\m}\epsilon_{\b}^{~\l\m_1}\bar{\nabla}_\l\Phi^{\b\m_2\m_3}.
\ee
Using the equations of motion, we get
\bea
\Pi_L^{(2)\m_1\m_2\m_3}&=&-\frac{\sqrt{-g}g^{00}}{64\pi G\m l}\Phi_L^{\m_1\m_2\m_3},\\
\Pi_R^{(2)\m_1\m_2\m_3}&=&\frac{\sqrt{-g}g^{00}}{64\pi G\m l}\Phi_R^{\m_1\m_2\m_3},\\
\Pi_M^{(2)\m_1\m_2\m_3}&=&-\frac{\sqrt{-g}g^{00}}{64\pi G}\Phi_M^{\m_1\m_2\m_3}.
\eea

The Hamiltonian is now
\be
H=\int d^2x (\dot{\Phi}_{\m_1\m_2\m_3}\Pi^{(1)\m_1\m_2\m_3}+\dot{K}_{\m_1\m_2\m_3}\Pi^{(2)\m_1\m_2\m_3}-{\cal L}_2).
\ee
Therefore, the energies of different spin-3 fluctuations are
\bea
E_M&=&\frac{2}{T\m}(\m^2-\frac{1}{l^2})\int d^3x \frac{\sqrt{-g}}{64\pi G}\epsilon_{\b}^{~0\m_1}\Phi_M^{\b\m_2\m_3}\dot{\Phi}_{M\m_1\m_2\m_3},\\
E_L&=&-\frac1T(1-\frac{1}{\mu l})\int d^3x \frac{\sqrt{-g}}{32\pi G}\bar{\nabla}^0\Phi_L^{\m_1\m_2\m_3}\dot{\Phi}_{L\m_1\m_2\m_3},\\
E_R&=&-\frac1T(1+\frac{1}{\mu l})\int d^3x \frac{\sqrt{-g}}{32\pi G}\bar{\nabla}^0\Phi_R^{\m_1\m_2\m_3}\dot{\Phi}_{R\m_1\m_2\m_3}.
\eea
Here the integration along $\tau$ direction is over $[0, T]$, and we choose $T=2\pi/(2\m l+1)$ for the massive mode and choose $T=2\pi/3$ for the left-moving and right-moving modes. Substituted $\Phi$ in the above equations by the real part of the solution in Eqs.~(\ref{solution1}-\ref{solution2}), we can show that the above three
integrals are negative for primary fields.\footnote{We use the same orientation as \cite{chiral}.} Then $\mu l>1$ gives $E_M<0$, while $\mu l<1$ gives $E_L<0$. So only at the critical point $\mu l=1$, there are no modes with negative energy, and in this case, we have $E_M=E_L=0$. By using the commutation relation of the generators, we can show the same result applies for the descendants. Therefore at the chiral point the left and the massive modes have zero energy, suggesting that they are just pure gauge.

\section{Conclusions and discussions}

In this paper, we studied the  spin-3 topologically massive gravity theory, especially at the chiral
point $\m l=1$. Inspired by the fact that first order formulation of TMG could be rewritten as a Chern-Simons gravity
plus a term imposing the torsion free condition, we proposed an action describing the spin-3 field coupled to 3D TMG.
Firstly we replaced the $SL(2,R)$ gauge group with $SL(3,R)$ group, and moreover imposed another torsion free condition
on spin-3 field. Next we rewrote the CS action in terms of the framelike field, and obtained the action (\ref{action}).
Starting from the action, we worked out the equations of motion. We showed that the fluctuations of spin-2 and traceless spin-3 fields around the AdS$_3$ vacuum with vanishing spin-3 field could be described by third order differential equations.  At generic value of $\m l$,
there exist a local massive degree of freedom for traceless spin-3 field, with conformal weight $(\m l+2, \m l-1)$. We found that at the chiral point
this massive mode become degenerate with the left-moving massless mode. We computed the energies of the fluctuations and showed that both the
massive mode and left-moving massless mode have zero energies at the chiral point, indicating that they are pure gauge. We also discussed the log modes of the fluctuation, which could be truncated out by imposing the generalized Brown-Henneaux boundary condition \cite{Henneaux:2010xg,theisen}. Therefore we obtained the same picture as the spin-2 fluctuations. In short, with the generalized Brown-Henneaux boundary conditions at the chiral point, there are only right-moving
boundary degrees of freedom, suggesting the boundary theory is chiral.

As the chiral gravity, the chiral spin-3 gravity could be described by
a holomorphic Chern-Simons gravity theory with the gauge group $SL(3,R)$ and the level $2k$.  The action (\ref{CS form}) at the chiral point $\m l=1$ reduces to a Chern-Simons term plus two terms relating to the torsions. If we take the vacuum to be AdS$_3$ with vanishing spin-3 field, the $\tilde \b^a$ and $\tilde \b^{ab}$ are both vanishing. But they do play a key role in studying the fluctuations around the vacuum and induce higher derivative terms. Nevertheless, once the fluctuations become pure gauge at the chiral point, it is safe to ignore these two terms and the action is just a holomorphic Chern-Simons gravity, from which the asymptotic symmetry could be read straightforwardly\cite{theisen, Banados:1998ta,Banados:1998gg}. It turns out to be   a classical $W_3$ algebra with central charge $c_R=3l/G$, with the generalized Brown-Henneaux boundary conditions. Namely the chiral spin-3 gravity could be described holographically by a two-dimensional chiral CFT with $W_3$ algebra and central charge $c_R$.

There are many interesting questions to be investigated in the future. We just list some of them:
\begin{enumerate}
\item In this paper, we focused the construction of the spin-3 field coupled to TMG. It would be interesting
to generalize the construction to arbitrary spin $N$. In particular, it would be nice to discuss its large N limit.
Alternatively, we may start from the higher spin algebra directly and construct the action, similar to the study in \cite{Henneaux:2010xg}.

\item We speculated that the chiral higher spin gravity is equivalent to a holomorphic Chern-Simons theory, indicating that the partition
function should be holomorphic. It would be nice to show this point explicitly;

\item It has been pointed out that in TMG there could exist another set of consistent boundary conditions, except the Brown-Henneaux boundary conditions, for the metric fluctuations. Imposing these boundary conditions may lead to another dual picture in terms of a logarithmic CFT\cite{Maloney:2009ck}. This is because that in TMG, the fluctuations in general satisfy a third order differential equations of motion, which allows the existence of log mode. Similarly in our theory, the spin-3 fluctuations allow a log mode as well. It is quite possible that there may
    exist other consistent boundary conditions for the spin-3 field, which leads to another CFT dual description.

\item It would be interesting to set up a large N duality for higher spin TMG, as the ones suggested for higher spin AdS$_3$ gravity\cite{Gaberdiel:2010pz,{Gaberdiel:2011wb}}.

\item  The BTZ black hole\cite{Banados:1992wn} is certainly the solution of the theory and its entropy in chiral gravity had been successfully produced from dual CFT. In our formulation, it would be nice to study the black hole solution with higher spin hair\cite{Gutperle:2011kf,{Ammon:2011nk}}.

\item It should be possible to generalize the construction to the supersymmetric TMG theory;

\item Most interestingly, from our point of view, our construction suggest that potentially the higher spin field theory
could be defined in warped spacetime, besides the usual AdS spacetime. Note that with vanishing spin-3 field, the equations of motion
from the action (\ref{action}) allow the solutions as the warped and null spacetimes and their related black holes\cite{Andy08,Bin10selfdual}. The spin-3 field could couple with these spacetime consistently via the action (\ref{action}). It is very interesting to study the spin-3 fluctuations
in these non-trivial backgrounds. Moreover, it would be valuable but very challenging to investigate the solutions with nontrivial
higher spin hair.

\end{enumerate}
We wish we can come back to these issues in the future.

{\bf Notes}: In our analysis of spin-3 fluctuation, we focused on its traceless part.
There is actually a trace part of spin-3 fluctuations, which has been studied in \cite{Bagchi11}. Such massive trace mode has zero energy and becomes pure gauge at
the chiral point. Therefore our conclusions on chiral spin-3 gravity are unaffected.

\noindent
 {\large{\bf Acknowledgments}}

 The work was in part supported by NSFC Grant No. 10975005. BC would like to
 thank the organizer and participants of the advanced workshop ``Dark Energy
 and Fundamental Theory" supported by the Special Fund for Theoretical Physics
 from the National Natural Science Foundations of China with grant no.: 10947203
 for stimulating discussions and comments.

\end{document}